          [1999/12/02 1.01 (PWD)]
\documentclass[a4paper,twocolumn]{esapub} 
\usepackage{natbib}
\usepackage{graphicx}

\title{Observations and theoretical models of gamma-ray bursts}
\author{A. J. Castro-Tirado}
\affil{Instituto de Astrof\'{\i}sica de Andaluc\'{\i}a (IAA-CSIC),
           Apdo. 03004, 18080, Granada, Spain}
\affil{Laboratorio de Astrof\'{\i}sica Espacial y F\'{\i}sica Fundamental 
           (LAEFF-INTA), Apdo. 50727, 28080, Madrid, Spain}

\begin{document}


\maketitle

\begin{abstract}
GRBs have remained a puzzle
for many high--energy astrophysicists since their discovery in 1967.
With the advent of the X--ray satellites {\it BeppoSAX} and {\it RossiXTE}, 
it has been possible to carry out deep multi-wavelength observations of 
the counterparts associated with the long GRBs class just within a few hours 
of occurence, thanks to the observation of the fading X-ray emission that 
follows the more energetic gamma-ray photons once the GRB event has ended. 
The fact that this emission (the afterglow) extends at longer wavelengths, 
has led to the discovery optical/IR/radio counterparts in 1997-2000, 
greatly improving our understanding of these sources.
Now it is widely accepted that GRBs originate at cosmological
distances with energie releases of 10$^{51}$--10$^{53}$ ergs. 
About 25 host galaxies have been detected so far for long-duration GRBs.
The observed afterglow satisfies the predictions of the 
"standard" relativistic fireball model, and the central engines that power 
these extraordinary events are thought to be the collapse of massive stars
rather than the merging of compact objects as previously also suggested. 
Short GRBs still remain a mystery as no counterparts have been detected.
\end{abstract}

\section{Introduction}

In 1967-73, the four VELA spacecraft (named after the spanish verb 
{\it velar}, to keep watch), that where originally designed for verifying 
whether the former Soviet Union abided by the Limited Nuclear Test Ban 
Treaty of 1963, observed 16 peculiarly strong events 
(Klebesadel, Strong and Olson 1973, Bonnell and Klebesadel 1996).
On the basis of 
arrival time differences, it was determined that they were related neither
to the Earth nor to the Sun, but they were of cosmic origin. Therefore they
were named cosmic Gamma-Ray Bursts (GRBs hereafter). Nearly 4000 events
have been detected to date. 

\section{Observational facts and implications}

\subsection{GRBs in the gamma-ray domain}

GRBs appear as brief flashes of cosmic high energy photons, carrying the 
bulk of their energy above $\approx$ 0.1 MeV. 
The KONUS instrument on {\it Venera 11} and {\it 12} gave the first 
indication that GRB sources were isotropically distributed in the sky 
(Mazets et al. 1981, Atteia et al. 1987).  Based on a much larger sample, this 
result was nicely confirmed by BATSE on board the {\it CGRO} satellite 
(Meegan et al. 1992). 
In general, there was no evidence of periodicity in the 
time histories of GRBs.  However there was indication of a bimodal 
distribution of burst durations, with
$\sim$25\% of bursts having durations around 0.2 s and $\sim$75\% with 
durations around 30 s. 
A deficiency of weak events was noticed in the log $N$-log $S$ diagram,
as the GRB distribution deviates from the -3/2 slope of the straight line 
expected
for an homogeneous distribution of sources assuming an Euclidean geometry. 
However, the GRB distance scale had to 
remain unknown for 30 years.  A comprenhensive review of these observational 
characteristics can be seen in Fishman and Meegan (1995).

\subsection{GRBs in the whole electromagnetic spectrum: eight selected bursts}

It was well known that an important clue for solving the GRB puzzle was going
to be the detection of transient emission -at longer wavelengths- associated 
with the bursts.
A review on the unsuccessful search for counterparts prior to 1997 
can be seen in Castro-Tirado (1998) and references therein.
Here I will present some results concerning eight selected bursts detected by 
the {\it BeppoSAX} ({\it BSAX}) and {\it RossiXTE} ({\it RXTE}) satellites in 
1996-2000 and their impact on the current understanding on the physics of GRBs.

\subsubsection{GRB 970228: the first X-ray and optical counterpart}

Thanks to {\it BSAX}, it was possible on 28 Feb 
1997 to detect the first {\it clear} evidence of a long X-ray tail 
 -the X-ray afterglow- following GRB 970228. A previously unknown 
X-ray source was seen to vary by a factor of 20 on a 3 days timescale. 
The X-ray fluence was $\sim$ 40 \% of the gamma-ray fluence, as reported 
by Costa et al. (1997), implying that the X-ray 
afterglow was not only the low-energy tail of the GRB, but also a significant 
channel of energy dissipation of the event on a completely different timescale.
Another important result was the non-thermal origin of the burst radiation 
and of the X-ray afterglow (Frontera et al. 1998). The precise X-ray 
position (1$^{\prime}$) led to the discovery of the first optical transient 
(or optical afterglow, OA) associated to a GRB,  
20 hr after the event with V $\sim$ 21.3 (Groot et al. 1997, van Paradijs 
et al. 1997). The OA was afterwards found on earlier images taken by 
Pedichini et al. (1998) and Guarnieri et al. (1997). The optical flux 
decayed following a power-law decay F $\propto$ t$^{-1.2}$ 
(Galama et al. 1997, Bartolini et al. 1998). 
An extended source was seen at the OA position since the very beginning 
by ground-based and {\it HST} observations (van Paradijs et 1997, 
Sahu et al. 1997).
New {\it HST} observations taken 6 months after the event were reported by 
Fruchter et al. (1997) and both the OA (at V = 28) and the extended source 
(V = 25.6) were seen. The latter object was interpreted as a 
galaxy, according to the similarities (apparent size, magnitude) with 
objects in the {\it HST} Deep Field. Finally, after two years, 
the redshift of this object has been determined as z = 0.695 (Djorgovski et
al. 1999a), confiming its extragalactic nature and implying a star-forming 
rate comparable to other galaxies at similar redshifts. 


\begin{figure}[t]
\centering 
\includegraphics[width=1.1\linewidth]{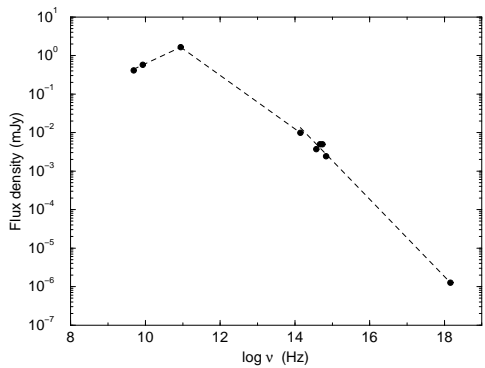} 
\caption{The multiwavelength spectrum of GRB 970508, on May 22, 1997.
         Adapted from Gorosabel (1999). See also Wijers and Galama 
         (1999).\label{fig:single}} 
\end{figure}

\subsubsection{GRB 970508: the clue to the distance}
The second OA associated to a GRB was discovered by Bond (1997) 
within the GRB 970508 error box, and observed 3 hr after the burst 
in unfiltered images (Pedersen et al. 1998). The optical light 
curve reached a peak in two days (R = 19.7, Castro-Tirado et al. 1998a, 
Djorgovski et al. 1997, Galama et al. 1998a) and 
was followed by a power-law decay F $\propto$ t$^{-1.2}$. 
Optical spectroscopy obtained during the maximum allowed a 
direct determination of a lower limit for the redshift of GRB 970805 
($z \geq 0.835$), implying E $\geq$ 7 $\times$ 10$^{51}$ erg 
and was the first proof that GRB
sources lie at cosmological distances (Metzger et al. 1997). The 
flattening of the decay in late August 1997 (Pedersen et al. 1998, Sokolov et
al. 1998) revealed the contribution of a constant brightness source 
-the host galaxy- that was revealed in late-time imaging obtained
in 1998 (Bloom et al. 1998, Castro-Tirado et al. 1998b, 
Zharikov et al. 1998).  
The maximum observed 1-day after the burst has not been detected in
other GRBs and it was interpreted by a delayed energy injection or
by an axially symmetric jet surrounded by a less energetic outflow
(Panaitescu et al. 1998).
The luminosity of the galaxy is well below the knee of the galaxy lumino\-sity 
function, L $\approx$ 0.12 $L^{*}$, and the detection of deep 
Mg I absorption (during the bursting episode) and strong [O II] 3727 $\rm \AA$ 
emission (the latter mainly arising in H II regions within the host galaxy) 
confirmed $z$ = 0.835 and suggested that the host could be a normal dwarf 
galaxy (Pian et al. 1998), with a star formation 
rate (SFR) of $\sim$ 1.0 $M_{\odot}$ year$^{-1}$ (Bloom et al. 1998).
Prompt VLA observations of the GRB 
970508 error box allowed detection of a variable radio source at 1.4, 4.8 
and 8.4 GHz, the first radiocounterpart ever found for a GRB (Frail et al. 
1997). The fluctuations could be the result of 
strong scattering by the irregularities in the ionized Galactic 
interstellar gas, with the damping of the fluctuations with time indicating 
that the source expanded to a significantly larger size.
However VLBI observations did not resolve the object (Taylor et al. 1997).
The transient was also detected at
15 GHz (Pooley and Green 1997) and as a continuum point source at 86 GHz with 
the IRAM PdBI on 19-21 May 1997 (Bremer et al. 1998).
A Fe K$\alpha$ line redshifted at $z$ = 0.835 in the X-ray 
afterglow spectrum (Piro et al. 1999) was attributed to a thick torus 
surrouding the central engine (M\'esz\'aros and Rees 1998).
GRB 970508 is the best observed afterglow so far. The broad band
spectrum (see Fig. 1) is nicely explained by the standard relativistic
blast wave model (Wijers and Galama 1999).

\subsubsection{GRB 970828: the first dark GRB}
This burst was detected by {\it RXTE} (Remillard et al. 1997) and was followed
up by {\it ASCA} and {\it ROSAT} (Murakami et al. 1997, Greiner et al. 1997).
 The fact that no optical counterpart down to R = 23.8 was detected between 
4 hr and 8 days after the event, could support the idea that the non-detection
was due to photoelectric absorption (Groot et al. 1998). The X-ray spectrum 
as seen by {\it ASCA} is strongly absorbed, suggesting that the event 
occurred in a dense medium. An excess at 6.7
keV was foud by {\it ASCA} in the X-ray afterglow spectrum. If this is due
to highly ionized Fe, then $z$ $\sim$ 0.33 (Yoshida et al. 1999) and the host
would be another dwarf galaxy (Gorosabel 1999). However, if the transient
radiosource detected with the VLA is indeed associated to the event, the
galaxy would be at z = 0.96 (Djorgovski et al. 2001), although there is
some concern (Murakami et al. 2001) regarding the proposed association as 
the rest frame energy of the line (9.79 keV) will be even larger than the
energy of the FeXXVII recombination edge at 9.28 keV (Weth et al. 2000).
 
And least in another three cases (GRB 981226, GRB 990506 and GRB 001109), 
radiotransients were detected without accompanying optical/IR transients. 
For GRB 000210, the {\it CHANDRA} position is consistent with a R = 23.5 
constant brightness object (Gorosabel et al. 2000). 

\subsubsection{GRB 980425: a GRB-SN connection ?}
A peculiar Type Ic supernova (SN 1998bw) was found in the error box for 
this soft GRB (Galama et al. 1998b). The SN lies in the galaxy ESO 184-G82, 
an actively star forming SBc sub-luminous galaxy at $z$ = 0.0085. 
The fact that the SN event occurred within $\pm$ 1 day 
of the GRB event, together with the relativistic expansion speed derived from 
the radio observation (Kulkarni et al. 1998a) strengths such a relationship. 
In that case, the total energy released would be 8 $\times$ 10$^{47}$ erg 
which is about $\sim$ 10$^{5}$ smaller than for "classical" GRBs.
Follow-up {\it HST} observations of ESO 184-G82 2.1 yr after the event, 
revealed an object 
consistent with being a point source within the astrometric uncertainty of 
0.018 arcseconds of the SN position. The object is located inside a star-
forming region and is at least one magnitude brighter than expected for the 
SN based on a simple radioactive decay model, implying either a significant
flattening of the light curve or a contribution from an underlying star 
cluster (Fynbo et al. 2000a). 

Reichart (1999) proposed a type Ib/c supernova lies "behind" another GRB 
(GRB 970228), overtaking the light curve two weeks after. This fact seems 
to be confirmed by the work of Galama et al. (2000). Castro-Tirado and 
Gorosabel (1999) and Bloom et al. (1999) also suggested the presence of
an underlying SN in GRB 980326 and recently in GRB 991208 (Castro-Tirado
et al. 2001).

\begin{figure}[t]
\centering 
\includegraphics[width=1.1\linewidth]{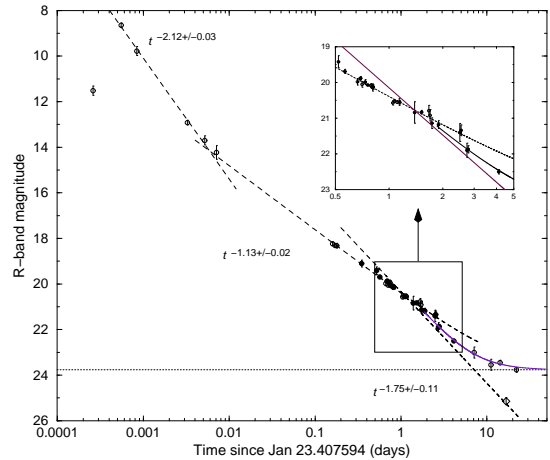} 
\caption{The R-band light-curve of the GRB~990123 OA.
 Based on our observations (filled circles) and other data reported elsewhere 
 (empty circles). 
 The doted line is the contribution of the  
 underlying galaxy, with R $\sim$ 23.77. 
 The three dashed lines are the contribution of the OA, following 
 F $\propto$ $t^{\delta}$ with $\delta$ = $-$2.12 up to $\sim$ 10 min, 
 $\delta$ = $-$1.13 up to $\sim$ 1.5 d, and
 $\delta$ = $-$1.75 after that time. The solid line, only drawn after 1.5 d 
 for clarity is the total observed flux (OA + galaxy).
 From Castro-Tirado et al. (1999).\label{fig:single}} 
\end{figure}

\begin{table*}[ht]
\hspace{1.5cm} 
\begin{center}
\caption{GRBs detected by {\it BeppoSAX/WFC} or {\it RXTE} in 1996-2000}
\begin{tabular}{lllllllll}
\hline
GRB    & X-rays   &opt-IR& radio & GRB    & X-rays   &opt-IR& radio \\
\hline
960720 &          &          &       & 990627 &  yes     &   no     &      \\
970111 &  yes ?   &   no     &  no   & 990704 &  yes     &          &      \\
970228 &  yes     &   yes    &  no   & 990705 &  yes     &   yes    &      \\
970402 &  yes     &   no     &       & 990712 &          &   yes    &      \\
970508 &  yes     &   yes    &  yes  & 990806 &  yes     &   no     &      \\
970616 &  yes ?   &   no     &  no   & 990907 &  yes ?   &   no     &      \\
970815 &  yes ?   &   no     &  no   & 990908 &          &   no     &      \\
970828 &  yes     &   no     &       & 991014 &  no      &   no     &      \\
971214 &  yes     &   yes    &  no   & 991105 &          &   no     &      \\
971227 &  yes ?   &   yes ?  &       & 991106 &  yes ?   &   no     &      \\
980109 &          &   yes ?  &       & 991216 &  yes     &   yes    & yes  \\
980326 &  yes ?   &   yes    &       & 991217 &          &          &      \\
980329 &  yes     &   yes    &  yes  & 000115 &  yes     &          &      \\
980425 &  yes ?   &   yes ?  & yes ? & 000210 &  yes     &   no     &      \\
980515 &  yes ?   &          &       & 000214 &  yes     &          &      \\
980519 &  yes     &   yes    & yes   & 000301c&          &   yes    & yes  \\
980613 &  yes     &   yes    &       & 000416 &          &          &      \\ 
980703 &  yes     &   yes    &  yes  & 000424 &          &   yes    & yes  \\ 
980706 &  yes ?   &   no     &       & 000508b&  yes     &          &      \\
981220 &  yes     &          &       & 000528 &  yes     &          &      \\
981226 &  yes     &          & yes ? & 000529 &  yes     &          &      \\ 
990123 &  yes     &   yes    & yes   & 000608 &          &          &      \\
990217 &  no      &   no     &       & 000615 &          &          &      \\
990308 &  yes ?   &   yes    &       & 000620 &  yes     &          &      \\
990506 &  yes     &   no     & yes ? & 001011 &          &   yes    &      \\
990510 &  yes     &   yes    &       & 001025 &  yes ?   &          &      \\
990520 &  yes     &   no     & no    & 001109 &  yes     &   no     & yes ?\\
990625 &          &          &       &        &          &          &      \\
\hline
\end{tabular} 
\end{center}
\end{table*}

\subsubsection{GRB 990123: the existence of a jet}
This is the first for which contemporaneous optical emission was
found simultaneous to the gamma-ray burst, reaching V $\sim$ 9  (Akerloff
et al. 1999). This optical flash did not track the gamma-rays and did not
fit the extrapolation of theSAX and BATSE spectra towards longer wavelengths. 
This optical emission was interpreted as the signature of a reverse shock
moving into the ejecta (Sari and Piran 1999). 
A brief radiotransient was also detected (Frail et al. 1999a)
coincident with the optical counterpart (Odewahn et al. 1999) and  
spectroscopy indicated a redshift z = 1.599 (Kulkarni et al. 1999, Andersen 
et al. 1999). A break observed in the light curve $\sim$ 1.5
days after the high energy event suggested the presence of a beamed outflow
(Castro-Tirado et al. 1999, Fruchter et al. 1999, Kulkarni et al. 1999). 
See Fig. 2. A weak magnetic field 
in the forward shock region could account for the observed multiwavelength 
spectrum in contrast to the high-field for GRB 970508 and it seems that
the emission from the three regions was first seen in this event (Galama et 
al. 1999a): the internal, reverse and forward shocks. 

\subsubsection{GRB 990510: first detection of linear polarization}
This burst belongs to the top 10\% of all GRBs detected by BATSE.
Following the BSAX/WFC detection, an optical counterpart was reported by 
Vreeswijk et al. (1999a) and $z$ = 1.619 was determined. 
The acromatic break seen in the light curve was
also interpreted as a jet, with a model yielding an opening angle of 
0.08 and a beaming factor of 300 (Harrison et al. 1999).
This is the first burst for which polarized 
optical emission was detected ($\Pi$ = 1.7 $\pm$ 0.2 \%), by means of an 
observation performed $\sim$18.5 hr after the event (Covino et al. 1999)
and later on (Wijers et al. 1999). 
This confirms the synchrotron origin of the blast wave itself and represents
the second case for a jet-like outflow (Stanek et al. 1999).
Further polarization measurements were carried out in GRB 990712 during one-
day time interval. The polarization angle did not vary significantly but the
degree of polarization was not constant. No current model can account for
this (Rol et al. 2001).

\begin{table*}[]
\hspace{1.5cm} 
\begin{center}
\caption{GRB host galaxies}
\begin{tabular}{llcl}
\hline
GRB    & R$_{host}$ &      $z$    &  References\\
\hline
970228 &  24.6    &         0.695        &  Djorgovski et al. (1999a)\\
970508 &  25.0    &         0.835        &  Bloom et al. (1998)      \\
970828 & 24.2 ?   &     0.33 ? 0.96 ?    &  Yoshida et al. (1999), Djorgovski et al. (2001a)   \\
971214 &  25.6    &         3.418        &  Kulkarni et al. (1998b)  \\
980326 &$\geq$27.3 &         1 ?         &  Bloom et al. (1999)      \\
980329 &  28.2    &                      &  Holland et al. (2000a)   \\
980425 & $\sim$15 &        0.0085        &  Galama et al. (1998), Fynbo et al. (2000a)    \\
980519 & $\sim$26 &                      &  Hjorth et al. (1999)     \\
980613 &  23.7    &         1.096        &  Djorgovski et al. (2001b)\\
980703 &  22.4    &         0.966        &  Djorgovski et al. (1998) \\
981226 &  24.9    &                      &  Frail et al. (1999b)     \\
990123 &  24.6    &         1.599        &  Kulkarni et al. (1999), Andersen et al. (1999)    \\
990506 &  24.8    &                      &  Frail et al. (2000)      \\
990510 &   28     &         1.619        &  Vreeswijk et al. (1999b), Bloom et al. (2001) \\
990705 &   23     &         0.85         &  Holland et al. (2000b), Amati et al. (2000) \\
990712 &  21.8    &         0.430        &  Galama et al. (1999b)    \\
991208 &  24.3    &         0.707        &  Dodonov et al. (1999), Castro-Tirado et al. (2001)    \\
991216 &  26.9    &         1.02         &  Vreeswijk et al. (1999c,2000) \\
000131 &$\geq$25.7 &        4.50         &  Andersen et al. (2000)   \\
000210 &   23.5   &                      &  Garmire et al. (2000), Gorosabel et al. (2000)    \\
000214 &          &         0.47 ?       &  Antonelli et al. (2000)  \\
000301c&$\geq$28.5 &        2.03         &  Smette et al. (2000,2001), Jensen et al. (2001)     \\
000418 &   24.0   &         1.118        &  Bloom et al. (2000)      \\
000926 &   23.9   &         2.066        &  Fynbo et al. (2000c,2001), Price et al. (2001)   \\
001109 &   20.7   &                      &  Taylor et al. (2000), Greiner et al. (2000)      \\
\hline
\end{tabular}
\end{center}
\end{table*}

\subsubsection{GRB 991216: a jewel for X-ray spectroscopy}
This GRB was detected by BATSE (Kippen et al. 1999) and subsequently 
scanned by {\it RXTE} detecting a bright  X-ray afterglow about 1 and 4-hr
after the occurence of the event (Takeshima et al. 1999, Corbet and 
Smith 1999). The optical and radio counterparts were identified by 
Uglesich et al. (1999) and Taylor et al. (1999) respectively. Three 
absorption systems at redshifts z = 0.77, 0.80 and 1.02 were found by 
Vreeswijk et al. (1999c) in the OA spectra.
A follow-up X-ray observation by {\it CHANDRA} has revealed the presence 
of a redshifted K-$\alpha$ line from H-like Fe (at rest energy of 6.97 keV).
See Fig. 3. The line width and intensity imply that the progenitor of the
GRB was a massive star system that ejected, shortly before the GRB event, 
about a $\sim$ 0.1 $M_{\odot}$ of Fe at 0.1$c$, probably by a SN explosion 
(Piro et al. 2000), at distances less than a light-hour 
(Rees and M\'esz\'aros 2000) thus giving support to the ``supranova'' model 
discussed in section 4.2 (Vietri et al. 2001).

\begin{figure}[t]
\centering 
\includegraphics[width=0.8\linewidth]{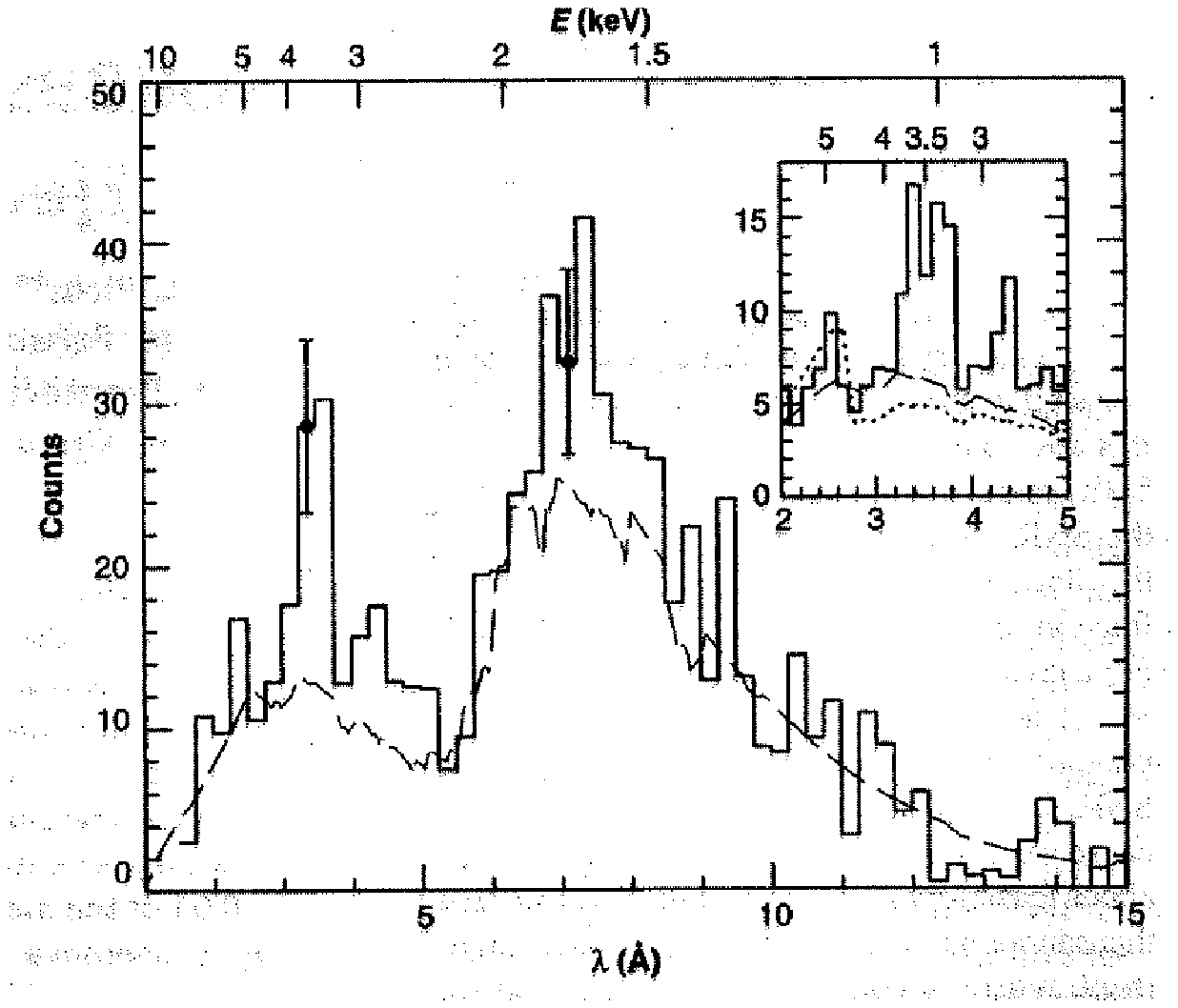} 
\caption{The X-ray spectrum of GRB~991216 as seen by CHANDRA HST on 17 
 Dec 1999. The dashed line is the best-fit power-law on the 0-th order
 ACIS-S spectrum. The narrow emission line at 3.5 keV (4.4 $\sigma$) is 
 identified as the redshifted K-$\alpha$ line from H-like Fe (at rest 
 energy of 6.97 keV). Further evidence is given by a recombination edge 
 in emission observed at 4.4 keV (with rest energy of 9.28 keV). 
 From Piro et al. (2000).\label{fig:single}} 
\end{figure}

\subsubsection{GRB 000301c: a very peculiar afterglow}
It was detected by {\it RXTE, Ulysses} and {\it ~Near}, and following
IPN coordinates (Hurley et al. 2000), a very blue optical counterpart was
found (Fynbo et al. 2000b, Jensen et al. 2001). 
The R- and K'-bands show significant differences, especially in the early
time decay slope and break time. This is contrary to the expectation of 
achromatic light curve breaks (Rhoads and Fruchter 2001). 
Therefore jet collimation is not the responsible mechanism
for the sharp break.
Refreshed shock effects can account to the high variability observed at optical
wavelengths (Masetti et al. 2000, Sagar et al. 2000), but also an ultra-
relativistic shock in a dense medium rapidly evolving to a non-relativistic 
phase (Dai and Lu 2001).
An UV spectrum with {\it HST} (+ NUV-MAMA) allowed to detect the H I Lyman-
break at $z$ = 2.03 (Smette et al. 2000, 2001).
A gravitational microlens has been also proposed to explain the short-time
scale variability (Garnavich, Loeb and Stanek 2000).

Further X-ray afterglows were observed by {\it BSAX} and {\it RXTE} in 
1997-2000, with exponents for the power-law decay in the X-rays and in the 
optical are in the range $\alpha$ = 0.75-2.25 for a dozen of bursts. Energies 
releases are of the order of 5 $\times$ 10$^{51}$ -- 2 $\times$ 10$^{54}$ erg.
These results are given on Table 1. See also Greiner (2001) for an updated 
information.

\section{GRB host galaxies}

About 50\% of the GRBs with X-ray counterparts are not detected in the 
optical, and this could be due to intrinsic faintness because of a low 
density medium, high absorption in a dusty enviroment, or Lyman limit 
absorption in high redshift galaxies (z $>$ 7).  
If GRBs are tightly related to star-formation, a substantial fraction of
them should occur in highly obscured regions. For instance, most of star 
formation in the Hubble Deep Field is so enshrouded by dust that starlight
from the galaxies detected by SCUBA is attenuated by a factor of 
$\sim$10$^{2}$ (Hughes and Dunlop 1999).
About 25 host galaxies have been detected so far, in the range 0.430 
$\leq$ z $\leq$ 4.50 if ESO 184-G82 is excluded. None of the hosts are 
brigther than the knee of the luminosity function L$^{*}$ at their 
redshift, but the GRB hosts are noticeable bluer than typical galaxies of 
similar magnitude (Fruchter et al. 1999).
Table 2 summarizes the properties of the host galaxies found so far.
See also Sokolov et al. (2001).

\section{Theoretical models}

\subsection{The standard model of GRB afterglows}

The observational characteristics of the GRB counterparts can be
accommodated in the framework of the relativistic fireball models, first 
proposed by Goodman (1986) and Paczy\'nski (1986), in which a compact source 
releases 10$^{53}$ ergs of energy within dozens of seconds in a region smaller
than 10 km. The opaque radiation-electron-positron plasma accelerates
to relativistic velocities (the fireball) with  Lorentz factors of 
$\Gamma$ $\sim$ 10$^{2}$--10$^{3}$. 
The GRB itself is thought to be
be produced by a serie of "internal shocks", at large radii probably within 
the fireball due to collisions amongst layers expelled with different
$\Gamma$  that are being caught up to each other (Rees and M\'esz\'aros 1994, 
Daigne and Mochkovitch 1998).
 
When the fireball runs into the surrounding medium, a "forward shock" 
ploughs into the medium, and sweeps up the interstellar matter, decelerating 
and producing an afterglow at frequencies gradually declining from X-rays 
to radio wavelenghts (M\'esz\'aros and Rees 1997). A "reverse 
shock" impinges on the ejecta. An extensive review is given by Piran (1999).
See Fig. 4.
Further collisions amongst faster shells and outer layers that have been 
decelerated are expected in the refreshed shock scenario where the ejecta
is reenergized. This can be done either continuosly or in discrete episodes
(Sari and M\'esz\'aros 2000).

The properties of the blast wave can be derived from the classical synchrotron
spectrum (Ginzburg and Syrovatskii 1965) produced by a population of 
electrons with the addition of self absorption and 
a cooling break (Sari, Piran and Narayan 1998).
The determination for every GRB of the six observables:
the synchrotron, break and self-absorption frequencies, the maximum flux and 
the power-law decay exponent (all from the multiwavelength spectrum) 
and $z$ (from optical or X-ray spectroscopy) allows to obtain 
the total energy per solid angle, the fraction of the shock energy in 
electrons and post-grb magnetic fields, and the density of the ambient medium.

\subsection{What are the progenitors ?} 
The most popular models fall into two broad cathegories: the explosion of 
a massive star and the coalescence of a compact binary system.

The ``collapsar'' model (Bodenhaimer and Woosley 1983, Woosley 1993, 2001) 
deals with a rotating massive star with a Fe core that collapses forming 
a Kerr black hole (BH) and a 0.1-1 $M_{\odot}$  torus. 
The matter is accreted at a very high rate and the energy can be extracted 
in two manners:
i)from the accretion of disk material by the BH; 
ii)from the rotational energy of the BH via the Blandford-Znajek process 
(Blandford and Znajek 1977), within the framework of a force-free magnetosphere
with a strong magnetic field and a magnetically dominated MHD flow 
(Brown et al. 2000, Spruit, Daigne and Drenkhahn 2001). The energy released
in this process is $\sim$ 10$^{54}$ erg.
A ``dirty fireball'', is produced 
reaching a luminosity $\sim$ 300 times larger that than of a normal SN. 
This would happen every $\sim$ 10$^{6}$ yr. 
In this scenario, GRBs would be produced in dense enviroments near star 
forming regions (see also MacFadyen and Woosley 1999) and GRBs might be used
for deriving the SFR in the Universe (Krumholz et al. 1998, Totani et al. 
1999).

Within this cathegory, the ``supranova'' model (Vietri and Stella 1998) 
involves a supra-massive neutron 
star imploding to a black hole and during the SN explosion the medium 
surrounding the remnant is swept-up, leading to a baryon-clean environment.
A torus of $\sim$ 0.1 $M_{\odot}$ is also expected and energy extraction
is via the conversion of the Poynting flux into a magnetized relativistic
wind. 

\begin{figure}[t]
\centering 
\includegraphics[width=0.9\linewidth]{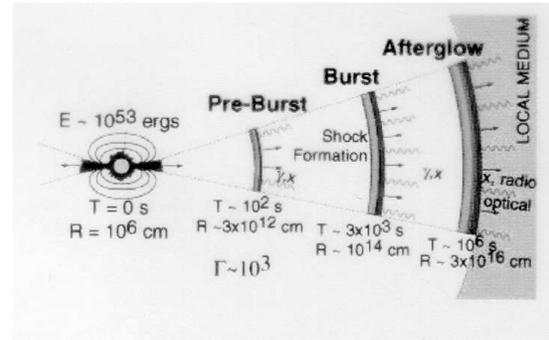} 
\caption{A scheme of the fireball plus relativistic blast wave model, 
 adapted from Piran (1999).\label{fig:single}} 
\end{figure}

The coalescence of neutron stars in a binary system has been also 
proposed (Narayan et al. 1992): lifetimes of such systems are of the order of
$\sim$ 10$^{9}$ years, and large escape velocities are usual, putting them 
far away from the regions where their progenitors were born. The likely 
result is a Kerr BH, and the energy released energy during 
the merger process is $\sim$ 10$^{54}$ erg. It is also possible that a
$\sim$ 0.1 $M_{\odot}$ accretion disk forms around the black hole and is 
accreted within a few dozen seconds, then producing internal shocks leading 
to the GRB (Katz 1997). 
In this scenario, GRBs would be produced far away from their host galaxies,
and this could account for the $\sim$ 40$\%$ of bursts not located in the 
optical window.

There are variations of this latter model where one or
two components are substituted for black holes (Paczy\'nski 1991), 
white dwarfs (Fryer and Woosley 1998) or He stars (Zhang and Fryer 2001).

A statitiscal study of the offsets of 20 long-duration GRBs from 
their apparent host galaxies centers (see Fig. 5) favours the explosion 
of a massive star 
rather than the binary merger model (Bloom et al. 2000).
It has been suggested that the short duration ($<$ 1 s) 
bursts could be due to compact star mergers, whereas the longer ones 
are caused by the collapse of massive stars.

\begin{figure}[t]
\centering 
\includegraphics[width=0.8\linewidth]{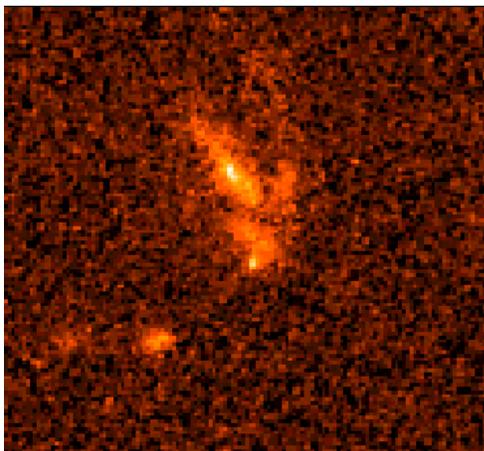} 
\caption{The host galaxy of GRB~990123 as imaged by HST on 23 March 1999. 
 The optical afterglow (the faint point source close to the centre of the
 image), which reached V = 8.9 simultaneously to the burst, faded to 
 V=27.7 by that time. While the whole galaxy and surroundings are relatively 
 blue, the OT does not fall on one of its bluest regions. 
 From Fruchter et al. (1999).\label{fig:single}} 
\end{figure}

\section{Summary}
The existence of X-ray afterglow in {\it most} bursts is confirmed.
Out ot 27 {\it BSAX} pointings, 18 revealed a clear afterglow, 
leading to the detection of several
optical/IR/radio counterparts in 1997-2000. 
However, only the population of bursts
with durations of few seconds has been explored. Short bursts lasting
less than 1 s, like GRB 980706, that follow the -3/2 slope in the 
log $N$-log $S$ diagram (in contrast to the longer bursts) remain to be 
detected at longer wavelengths.
In any case, it is clear today that GRBs might provide an important clue
for the study of the early Universe (Lamb \& Reichart 2000).

\end{document}